\documentclass[10pt,a4paper,twocolumn]{article}
\usepackage[utf8]{inputenc}
\usepackage[english]{babel}
\usepackage{url}
\usepackage[T1]{fontenc}
\usepackage{amsmath}
\usepackage{amssymb}
\usepackage{placeins}
\usepackage{graphicx}

\usepackage[font=small,labelfont={bf},figurename=Figure]{caption}

\usepackage[left=2cm,right=2cm,top=2cm,bottom=2cm]{geometry}

\usepackage[affil-it]{authblk}

\usepackage{appendix}

\usepackage[style=numeric-comp, sorting=none,firstinits=true,doi=false,isbn=false,url=false,eprint=false,maxbibnames=99]{biblatex}
\newcommand{\wn}{$\text{cm}^{-1}$}
\newcommand{\mic}{$\mu\text{m}$}
\newcommand{\mics}{$\mu\text{m}^2$}

\title{Ultra-fast amplitude modulation of mid-IR free-space beams at room-temperature}

\author[1,*]{Stefano Pirotta}
\author[1]{Ngoc-Linh Tran}
\author[2]{Giorgio Biasiol}
\author[1]{Arnaud Jollivet}
\author[1]{Paul Crozat}
\author[1]{Jean-Michel Manceau}
\author[1]{Adel Bousseksou}
\author[1,$\dagger$]{Raffaele Colombelli}

\affil[1]{Centre de Nanosciences et de Nanotechnologies (C2N), CNRS UMR 9001, Université Paris-Sud, Université Paris-Saclay, 91120 Palaiseau, France}
\affil[2]{Laboratorio TASC, CNR-IOM, Area Science Park, Basovizza, I-34149 Trieste, Italy}
\affil[*]{E-mail: stefano.pirotta@u-psud.fr}
\affil[$\dagger$]{E-mail: raffaele.colombelli@u-psud.fr}

\date{}

\begin{document}

\twocolumn[
\maketitle
\begin{@twocolumnfalse}
\begin{abstract}
Applications relying on mid-infrared radiation (Mid-IR, $\lambda\sim$ 3-30 \mic) have progressed at a 		very rapid pace in recent years, stimulated by scientific and technological breakthroughs. Mid-IR cameras
have propelled the field of thermal imaging. And the invention of the quantum cascade laser (QCL)
has been a milestone, making compact, semiconductor-based mid-IR lasers available to a vast
range of applications. All the recent breakthrough advances stemmed from the development of a
transformative technology. In addition to the generation and detection of light, a key functionality
for most photonics systems is the electrical control of the amplitude and/or phase of an optical
beam at ultra-fast rates (GHz or more). However, standalone, broadband, integrated modulators
are missing from the toolbox of present mid-IR photonics integrated circuits and systems
developers. We have developed a free-space amplitude modulator for mid-IR radiation ($\lambda\sim$ 10 \mic) that can operate up to at least 1.5 GHz (-3dB cut-off at $\sim$ 750 MHz) and at room-temperature. The device relies on a semiconductor hetero-structure enclosed in a judiciously designed metal-metal
optical resonator. At zero bias, it operates in the strong light-matter coupling regime up to 300K.
By applying an appropriate bias, the device transitions to the weak coupling regime. The large
change in reflectivity due to the disappearance of the polaritonic states is exploited to modulate
the intensity of a mid-IR continuous-wave laser up to speeds of more than 1.5 GHz.
\vspace{3em}
\end{abstract}
\end{@twocolumnfalse}]

\begin{refsection}[ModulatorMIR.bib]
\section*{Introduction}
\begin{figure*}[ht]
\centering
\includegraphics[width=0.9\linewidth]{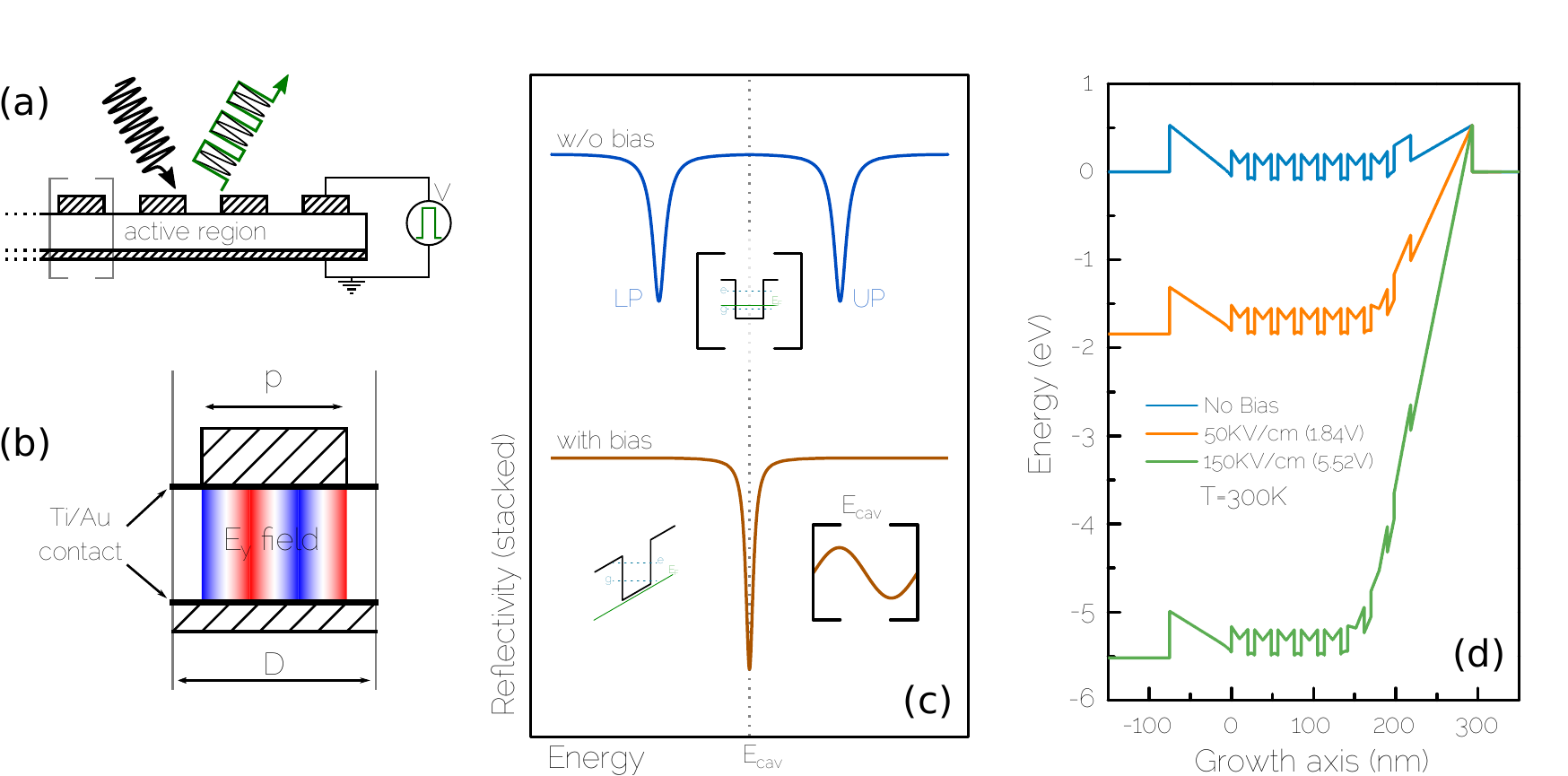}
\caption{ (a) Sketch of the modulator geometry: the active region is embedded in a metal-metal
structure. By applying on it an external bias, the amplitude of the reflected beam is
modulated. (b) Close-up of a single 1D ribbon of the device. p is the patch side, D the period. Two
Schottky contacts permit the effective application of an external bias to the hetero-
structure. The electric field distribution (y-component) of the TM$_{03}$ mode is also sketched.
TM$_{0i}$ refers to the mode with i nodal lines in the transverse electric field distribution.
(c) Intuitive view of the modulator operating principle: with no applied applied the system is
designed to be in strong coupling. Two polaritonic branches are visible in the reflectivity
spectrum, for a specific value of the patch side p. By applying a specific bias we can de-
populate an arbitrary number of quantum wells and bring the system in the weak-coupling
regime: the bare cavity TM$_{03}$ mode is visible on the reflectivity spectrum.
(d) Conduction band profile obtained by solving the Schrodinger-Poisson equation
(NextNano) at room temperature with increasing external bias. No bias (solid blue line), 1.84
V (orange solid line) and 5.52 V (green solid line).}
\label{fig:fig1}
\end{figure*}
Fast amplitude and phase modulation are essential for a plethora of applications in mid-IR
photonics, including laser amplitude/frequency stabilization \cite{Bernard1997}, coherent detection, FM and AM
spectroscopy and sensing, mode-locking, and optical communications \cite{Martini2001, Chuanwei2015}. However, the ultra-fast (1-40
GHz) modulation of mid-IR radiation is a largely under-developed functionality. The fastest
modulation speeds, 20-30 GHz, have been obtained with the direct modulation of mid-IR QCLs, but
this requires specially designed devices and elevated injected RF (radiofrequency) powers \cite{Paiella2001, Hinkov2019, Mottaghizadeh2017}.
Interestingly, in the visible/near-IR spectral ranges the preferred solution is to separate the
functionalities: independent modulators, filters, interferometers are employed that are physically
separated from the source. For modulators, this leads to advantages in terms of RF power, laser
linewidth and flatness of the modulation bandwidth.

Commercially available mid-IR modulators are either acousto-optic devices with narrow modulation
bandwidth, or very narrow band ($\sim$100 kHz) electro-optic modulators based on GaAs or CdTe \cite{Qubig2020}. The
latter ones can operate up to modulation speeds of 20 GHz, but their efficiency is very low (less than
0.1\% sideband/carrier ratio). To date, standalone, efficient and broadband amplitude/phase
modulators are missing from present mid-IR photonics tools.

One way forward is photonic integration: scaling to the mid-IR the approach already developed for
silicon photonics. It can rely on SiGe/Si photonic platforms \cite{Vakarin2017}, or on the more natural InGaAs/AlInAs-
on-InP platform \cite{Jung2019}. In both cases the QCL source must be properly integrated in the system.
An alternative is to develop modulators that can apply an ultra-fast RF modulation to a propagating
beam, either in reflection or in transmission. This approach does not require a specific integration of
the source and can in principle be applied to laser sources beyond QCLs.

In this letter we follow the latter strategy, and we demonstrate ultra-fast amplitude modulators of
mid-IR free-space beams that operate at room-temperature in a reflection configuration (scheme in
Figure \ref{fig:fig1}(a)) up to more than 1 GHz modulation speed.

Since the eighties, proposals have been put forward to exploit intersubband (ISB) absorption in
semiconductor quantum well (QW) systems to modulate mid-IR radiation. The first attempts based
on the Stark shift were then followed by a number of works exploiting coupled QWs \cite{Vodjdani1991,Dupont1993}. In both
cases, the application of an external bias depopulates the ground state of QW at 78K thus inducing a
modulation of ISB absorption \cite{Duboz1997}. Operation at room temperature was obtained in \cite{Berger1996} using a
Schottky contact scheme. Recently, different approaches have been proposed to actively tune the
reflectivity/transmission of mid-IR and/or THz beams: phase transition in materials like VO$_2$ , liquid
crystals orientation, carrier density control in metal-insulator-semiconductor junctions \cite{Jun2012}. These
devices operate on the principle that, at a given wavelength, a change in absorption translates in a
modulation of the transmitted power. An alternative approach is to frequency shift the ISB
absorption, instead of modulating its intensity. In \cite{Benz2013,Lee2014} a giant confined stark effect in a coupled
QW system embedded in a metallic resonator, designed to be in the strong coupling regime between
light and matter, was exploited. A response time of $\sim$ 10 ns was estimated.

\section*{Switching between strong and weak coupling regimes}
Our approach is to operate the device in the strong light-matter coupling regime, and introduce the
ultra-fast modulation by switching the system in and out of strong coupling with the application of
a bias voltage. A periodic QW structure is embedded in an optical resonator composed by non-
dispersive metal-metal 1D ribbons (or 1D patch cavities \cite{Todorov2010}) as shown in Figure \ref{fig:fig1}(a). The system,
designed to operate in reflectance, is conveniently optimized so that the ISB transition is strongly
coupled to the TM$_{03}$ photonic mode of the resonator, whose electric-field distribution is shown in Figure \ref{fig:fig1}(b). (The notation convention TM$_{0i}$ is defined in the caption). The resulting reflectivity R is sketched in
Figure \ref{fig:fig1}(c). At zero bias, at the cavity frequency $\nu_{cav}$ for instance, R$\sim$1 as the eigenmodes of the system
are the upper (UP) and lower (LP) polaritons. An applied bias transitions the system, totally or
partially, to the weak coupling regime, with a reflectivity drop down to R$\sim$0. A laser with frequency
$\sim\nu_{\text{cav}}$ (and so at $\nu_{\text{LP}}$ and $\nu_{\text{UP}}$) will be amplitude-modulated with an elevated contrast.

As the intersubband polariton dynamics features ps-level timescales \cite{Gunter2009}, the bandwidth of the
modulator is limited by (i) the RC-constant of the circuit and (ii) the transfer time of electrons in/out
of the QWs. In fact, the top and bottom metal-semiconductor Schottky interfaces permit the
application of a gate to the multiple QW structure that can efficiently deplete the system, as shown
in Figure \ref{fig:fig1}(d). Note: the electrical control of ISB polaritons, in a quasi-DC regime though, has been
studied in \cite{Anappara2005, Anappara2006}.

\begin{figure}[!th]
\centering
\includegraphics[width=\linewidth]{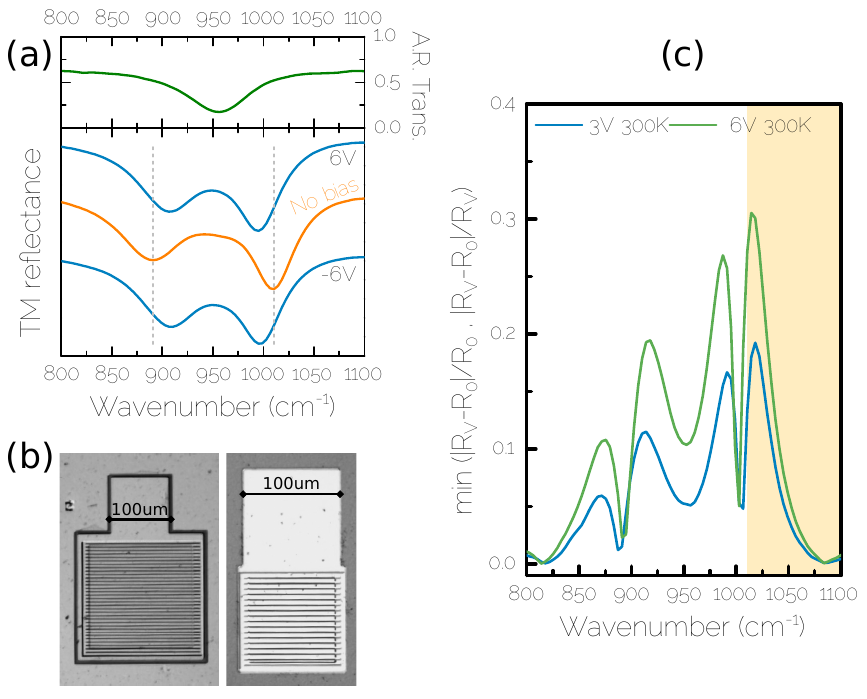}
\caption{(a) Top panel: transmission through the bare active region in a typical multi-pass geometry,
at room temperature: a clear ISB transition at about 955 \wn\ is measured.
Bottom panel: room-temperature reflectivity spectra under FTIR-coupled microscope of the
p = 4.2 \mic\ sample. The orange solid line is the reflectivity at no bias. The blue solid lines
corrspond to the application of $\pm$6V. The curves are stacked for clarity. The Rabi-splitting
decreases by 25\%. (b) Optical microscope images of the large devices (left; surface $5\times 10^4$ \mics) and of the small
ones (right; surface $2\times 10^4$ \mics). In both samples the bonding pad has the same dimensions
(100$\times$100 \mics). (c) Modulation depth extracted from the spectra of panel (b) (large devices). It is defined as $\min\left(\lvert\frac{R_{0V}-R_{6V}}{R_{6V}}\rvert\, , \, \lvert\frac{R_{0V}-R_{6V}}{R_{0V}}\rvert  \right)$ and it is plotted in the range 800-1100 \wn\ for both 3V (green solid
line) and 6V (blue solid line) biases. The semi-transparent orange region corresponds to the
wavelength coverage of our commercial, tunable mid-IR QC laser source.}
\label{fig:fig2}
\end{figure}

\section*{Sample fabrication}
The semiconductor hetero-structure was grown by solid-source molecular beam epitaxy on an un-
doped GaAs substrate. It is composed of 7 periods of 8.3 nm GaAs QWs separated by 20 nm-thick
Al$ _{0.33}$Ga$_{0.67}$As barriers. Si delta-doping ($\text{n}_{\text{Si}} = 1.74\times 10^{12} cm^{-2}$) is introduced in the barrier center. 
A 40 nm-thick GaAs cap layer terminates the structure, and a 500 nm-thick Al$_{0.50}$Ga$_{0.50}$As is introduced before the active region, 
whose total thickness is L$_{AR}$ = 368.1 nm. The sample presents an ISB transition at an
energy of 118.5 meV (about 955.8 \wn), that we have measured at 300K in a classic multipass waveguide
transmission configuration (green spectrum in Figure \ref{fig:fig2}(a) ). Figure \ref{fig:fig1}(d) shows the global conduction band
profile at room-temperature (RT, solid lines) and at different applied biases for the fabricated
structure. It was obtained solving self-consistently the Schrodinger-Poisson equations using a
commercial software \cite{Birner2007}. With no applied bias all the QWs are populated. The application of a bias
gradually depletes them.

The modulators rely on a metal-semiconductor-metal geometry. We have wafer-bonded the
sample on a n$^+$- GaAs carrier layer via Au-Au thermo-compression wafer-bonding, a standard
technology for mid-IR polaritonic devices \cite{Vigneron2019, Manceau2018}. After polishing and substrate removal, the 1D patches
are defined with electron-beam lithography followed by Ti/Au deposition (5/80 nm) and lift-off. The
top contact patterning and the definition of the bonding pads are realized with optical contact
lithography and Ti/Au lift-off. An inductively coupled plasma (ICP) etching step down to the back
metal plane defines the mesa structure. Optical microscope images of typical final devices are shown
in Figure \ref{fig:fig2}(b).
Arrays of devices have been fabricated that differ in the width p of the metallic fingers
(nomenclature in Figure \ref{fig:fig1}(b)). For each value of p, we fabricated two arrays with different total surface
($5\times 10^4$ and $2\times 10^4$ \mics , respectively. Figure \ref{fig:fig2}(b)). The active region being very thin (368.1 nm), the system
does not operate as a photonic-crystal, but operates instead in the independent resonator regime.
The cavity resonant frequency $\nu_{\text{cav}}$ is set by p, not by the period D, according to the following
expression:
\begin{equation*}
\frac{c}{\nu}=\lambda=\frac{2\, n_{\text{eff}}\, p}{i} \;\;  \text{where} \: i\in\mathbb{N} \: .
\end{equation*}
The system behaves as a Fabry-Perot cavity of length p, with n$_\text{eff}$ an effective index that takes into
account the reflectivity phase at the metallic boundaries \cite{Todorov2010, Duperron2013}. We opted to operate not on the i=1
fundamental mode, the standard choice \cite{Todorov2010}, but on the i=3 mode (the TM$_{03}$), to simplify the fabrication
procedure and increase the electromagnetic overlap factor. Figures \ref{fig:figS1} and \ref{fig:figS2} in Supplementary
Material provides a justification for this choice.

\section*{Experimental results}
\begin{figure}[ph]
\centering
\includegraphics[width=\linewidth]{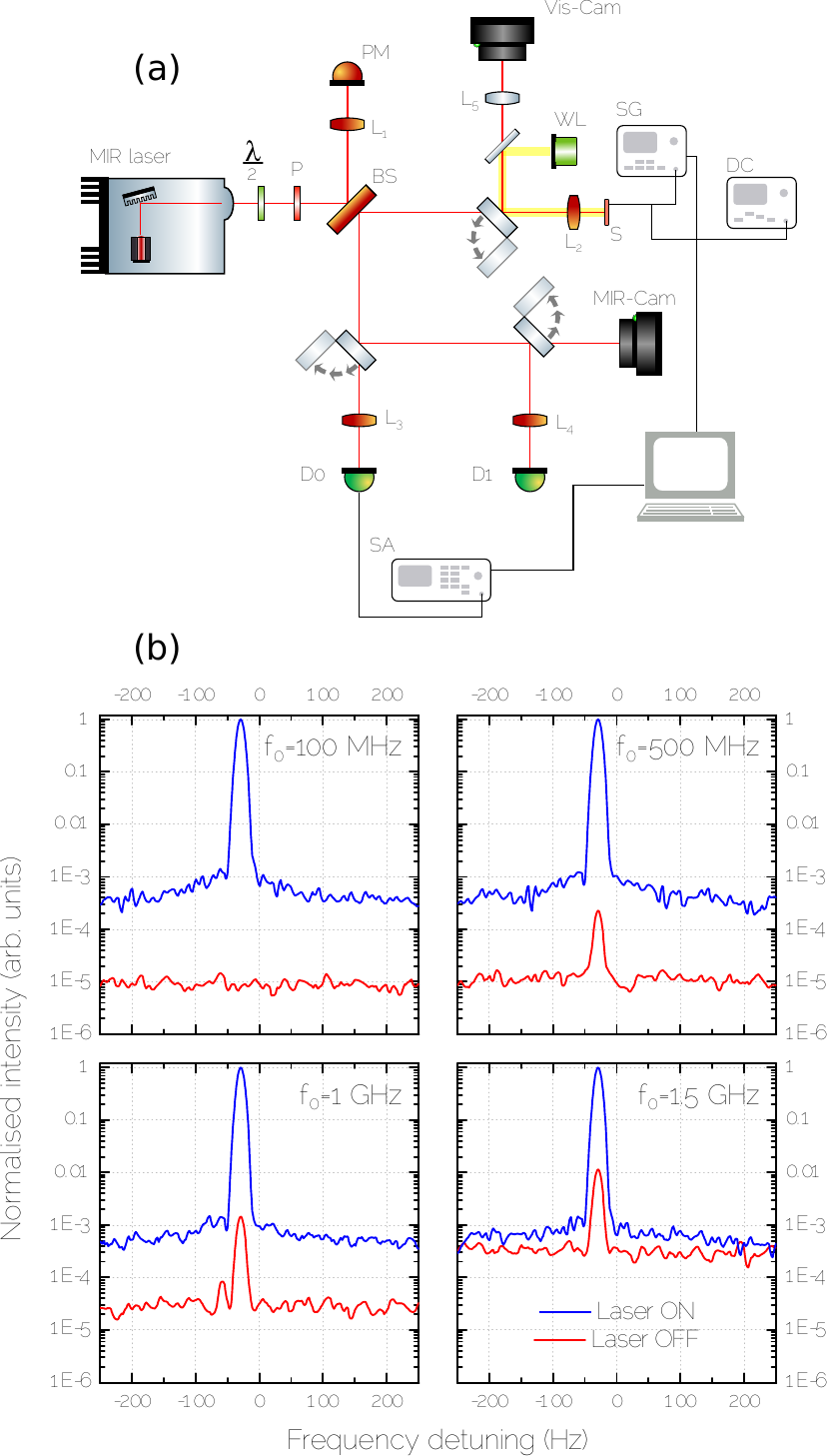}
\caption{(a) Sketch of the experimental setup to measure the modulator bandwidth. The sample S is
pumped with a commercial tunable mid-infrared QC laser focused with a ZnSe (L$_2$ ) lens; the
back reflected beam is collected through the beam-splitter BS, and sent to the
detectors D$_0$ (fast MCT Vigo detector with nominal bandwidth 1KHz - 837MHz) and D$_1$ (general
purpose 50 MHz bandwidth MCT detector, LN-cooled). A signal generator (RF or LF
depending on the measurement) is used to apply the electrical bias to the sample.
The electrical signal from the detector is then sent to specific analyzers: a spectrum analyzer
(SA) to collect the beat-note spectrum or a lock-in. PM: power meter; WL: white light; P: polarizer; $\lambda$/2: half-wave plate; Vis-Cam:
visible camera; MIR-Cam: MIR camera. (b) Normalized beat-note spectra obtained when the sample is fed with 10dBm/DC@-
989mV with modulation frequencies 100 MHz, 500 MHz, 1 GHz and 1.5 Ghz from top left.
The measurements are performed at room-temperature, and the measured sample is the
small one (surface is $2\times 10^4$ \mics ). The QC laser frequency is \wn. Two curves are shown
for each panel: laser on (solid blue line) and laser off (solid red line). The modulator performs
up to at least 1.5 GHz.}
\label{fig:fig3}
\end{figure}
The reflectance of the devices as a function of p has been measured with a microscope coupled to
an FTIR spectrometer to retrieve the polaritonic positions at RT (and at 78K) with no applied bias. The
complete dispersion is shown in Figure \ref{fig:figS2} : together with the
measurements and simulations on an empty cavity (Figure \ref{fig:figS1}) it permits to identify the first 3 ribbon
resonator modes. The TM$_{03}$ mode exhibits a clear Rabi splitting for patch sizes around p = 4 \mic.

A suitable device (p = 4.2 \mic) was wire bonded and its reflectance was measured under different
applied DC biases. When no bias is applied, we observe (orange curve in Figure \ref{fig:fig2}(a)) the 2 polariton
branches. The two blue solid lines correspond to a $\pm$6V applied bias, that is practically the limit
imposed by the Ti/Au Schottky barriers. The Rabi decreases by 25\%: it means that the gate empties
only half of the QWs, as $\Omega_{\text{Rabi}} \propto \sqrt{N_{\text{QW}}}$, being N$_{\text{QW}}$ 
the total number of QWs in the structure. (A different, lower doped sample that fully transitions to
the weak coupling regime is reported in Figure \ref{fig:figS3} of Suppl. Material).

From the measurements we can extract the modulation depth attainable on an incoming laser
beam with a +6V maximum bias with this specific device. The modulation is defined as
$\min\left(\lvert\frac{R_{0V}-R_{6V}}{R_{6V}}\rvert\, , \, \lvert\frac{R_{0V}-R_{6V}}{R_{0V}}\rvert  \right)$
and it is plotted in Figure \ref{fig:fig2}(c) in the 800 - 1100 \wn\ range. It shows that a contrast above 10\% can be obtained 
in a few frequency ranges. In particular, a contrast between 20\% and 30\% can be obtained around 1030 \wn\ ($\lambda\sim$ 9.70 \mic). 
This frequency is covered by our tunable commercial QC laser (shadowed orange region in Figure \ref{fig:fig2}(c)).

We have measured the speed and modulation bandwidth of the modulator with the setup
described in Figure \ref{fig:fig3}(a). A continuous-wave (CW), tunable commercial QC laser \cite{Day2020} is
focused on the modulator (S) that is fed with an RF signal from a synthesizer (SG). The reflected, and
modulated, beam is detected with a 837 MHz-bandwidth commercial MCT detector (D0, \cite{Vigo2020}) whose output is
fed to a spectrum analyzer (SA), or – for low frequency measurements – to a 50 MHz-bandwidht MCT detector (D1)
and then to a 200 MHz lock-in amplifier. All measurements are performed at room temperature
(300K).

\begin{figure*}[h!]
\centering
\includegraphics[width=0.8\linewidth]{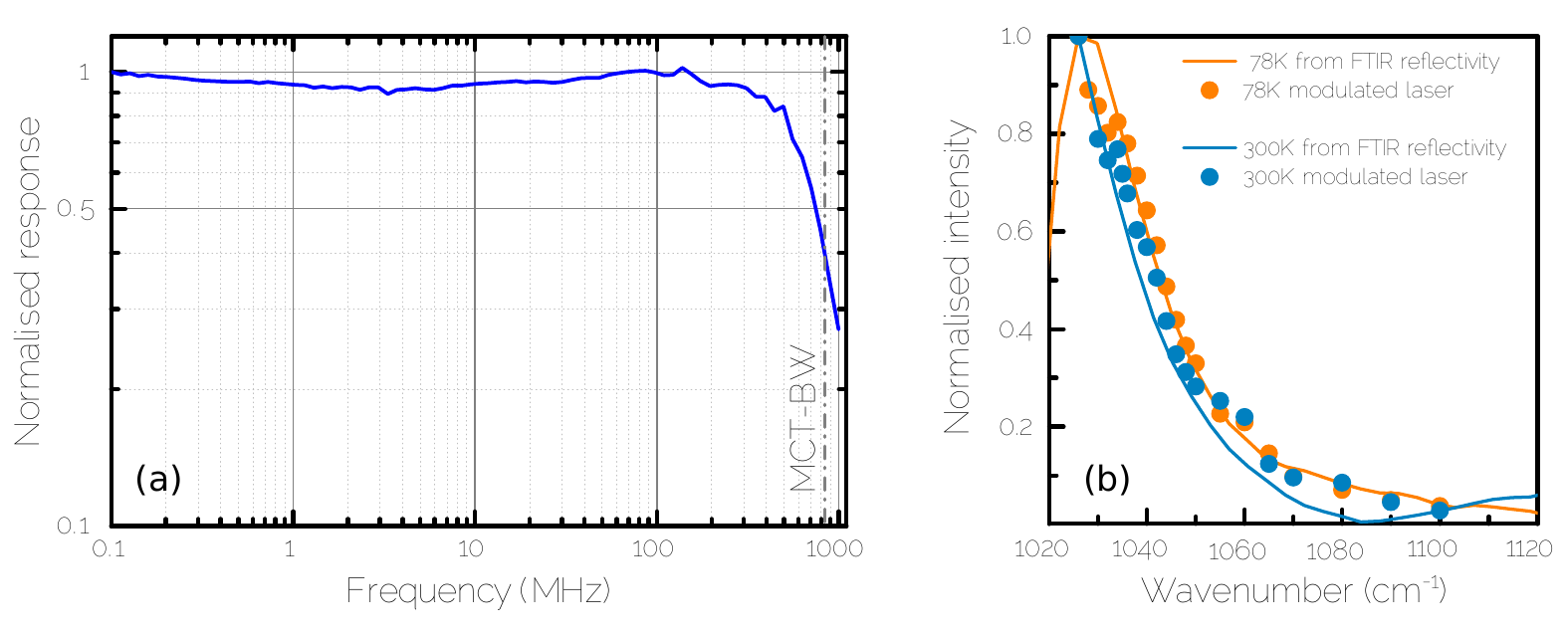}
\caption{(a) Normalized response at room temperature of the small device when fed with 10dBm/
DC@-989mV. The QC laser frequency is 1010 cm -1 . The nominal -3dB cut-off at 837MHz of
the fast MCT is shown as a grey dashed vertical line. The entire experiment is automated by
an homemade python code, based on the Pymeasure package \cite{PyMeasure2020}. (b) Normalized modulation depth as a function of the QC laser frequency (dots). The
measurements are performed on the large devices (surface is $5\times 10^4$). The QC laser
frequency is tuned between 1026 and 1100 \wn\ and the modulated beam is detected by the
50 MHz MCT and the lock-in. The modulator is fed with a 0 - 3 V sinusoidal signal at 10 kHz.
The solid line is the modulation depth of Figure \ref{fig:fig2}(c), normalized to 1 at 1026 \wn . The perfect
overlap proves that the origin of the modulation is indeed the transition between strong and
weak light-matter coupling regimes.}
\label{fig:fig4}
\end{figure*}

Figure \ref{fig:fig3}(b) shows the spectra obtained from the smaller modulator (p=4.1 \mic), using a QCL
frequency of 1010 \wn. We can detect a signal up to a modulation speed of 1.5 GHz, well beyond the
VIGO detector 3dB cutoff of 837 MHz, proving the ultra-fast character of our modulator. In order to
determine the modulator bandwidth, we performed an automated scan as a function of the
modulation frequency. 
The results, at 300K, are reported in Figure \ref{fig:fig4}(a) for a typical 100x100 \mic$^2$\ large  device, with grating period p=4.1 \mic.
At the optimum performance point ($\nu_\text{laser}$ = 1010 \wn), it operates at
frequencies > 1GHz, with a -3dB cut-off at $\sim$ 750 MHz (Figure \ref{fig:fig4}(a), blue curve). 
The larger devices (data not shown) typically exhibit a -3dB cut-off at $\sim$ 150 MHz.
This result is in fair agreement with the surface ratio between the two devices.
Furthermore the theoretical RC-cutoff of the large samples is $f^\text{large}_\text{cutoff}=\frac{1}{2\pi RC}\,=$ 204MHz and for the small sample $f^\text{small}_\text{cutoff}\,=$ 510MHz (C device capacitance and R the 50 $\Omega$ output resistance of the RF syntesizer). The good agreement
proves that the bandwidth is currently limited by the RC time constant. For high-resolution
spectroscopy, an important parameter is the sideband/carrier power ratio. For the current
modulators, we estimate a ratio of the order of 5\%.

If the QC laser frequency is tuned away from the optimum value, Figure \ref{fig:fig2}(c) predicts that the
modulation contrast should drop. This observation is crucial to unambiguously assign to the polariton
modulation the enabling physical principle of the device. To this scope we have measured the
modulation contrast as a function of the QCL laser frequency, at both RT and 78K. The results –
normalized to 1 - are reported in Figure \ref{fig:fig4}(b) (dots) for the larger sample, and they are superimposed to
the DC modulation response curve obtained from the reflectivity measurements from \ref{fig:fig2}(c). The
modulator ultra-fast response as a function of the impinging laser frequency closely follows the DC
contrast curve. This finding confirms that the enabling mechanism is indeed the ultra-fast modulation
of the Rabi splitting \textit{via} application of an RF signal.

\section*{Discussion and conclusions}
Having established that the current devices are RC limited, the natural question is: what is their
intrinsic speed? The physics of the current devices is not very different from the one of mid-IR QWIP
detectors, except the absence of the ohmic contacts, that are known to operate up to speeds of
60/80 GHz \cite{Schneider2007,Lin2019}. That alone suggests that the intrinsic speed of the current modulators is set by the
same parameters, in particular the capture time $\tau_{cap}$ and the transit time $\tau_{trans}$ that is set by the drift
velocity. There is however a notable difference: in the ideal operating regime, carriers diffuse all the
way towards one metal-semiconductor interface upon application of a bias. And they have to flow
back through the active region when the bias is restored to 0. This leads of a characteristic time $\tau_\text{drift}=\frac{L_\text{AR}}{v_\text{drift}}$. 
In our case, with a very conservative $v_\text{drift}=10^6\,\frac{\text{cm}}{\text{s}}$ at RT \cite{Hava1993}, we obtain a value of 
$\tau_\text{drift}<30$ ps which sets a lower bound for the intrinsic cut-off at < 10 GHz.

In conclusion, we have demonstrated a technology that is able to amplitude modulate mid-IR free
space laser beams up to GHz modulation frequencies. In this first demonstration, at $\lambda$ = 9.7 \mic, we
achieved modulation speeds up to 1.5 GHz at room temperature (-3dB cut-off at $\sim$ 750 MHz). The
device operates by transitioning in and out of the strong coupling regime at ultra-fast rates, and it is
possibly the first report of a practical device relying on the strong light-matter coupling regime. The
estimated intrinsic speed is at minimum 10 GHz. Improved active regions that do not rely on drift
transport, but instead on tunnel coupling \cite{Anappara2006} will probably lead to modulation speed in the 30/40 GHz
range.

\section*{Funding Information}
We acknowledge financial support from the European Union FET-Open Grant MIRBOSE (737017). This
work was partly supported by the French RENATECH network. R.C. and A.B. acknowledge financial
support from the French National Research Agency (project “IRENA”). 

\section*{Acknowledgments}
We thank S. Barbieri, J-F. Lampin and E. Peytavit for useful discussions. We also thank L. Wojszvzyk, A.
Nguyen and J-J. Greffet for the loan of the 50 MHz-bandwidth MCT detector.\\

\noindent\textbf{Disclosures.} The authors declare no conflicts of interest.\\

\noindent See Supplementary material below for supporting content.

\printbibliography
\end{refsection}

\newpage
\clearpage
\onecolumn

\begin{refsection}[ModulatorMIR_supp.bib]
\appendixpage
\section*{Bare cavity reflectivity}
\renewcommand{\thefigure}{S1}
\begin{figure}[ht]
\centering
\includegraphics[width=0.5\linewidth]{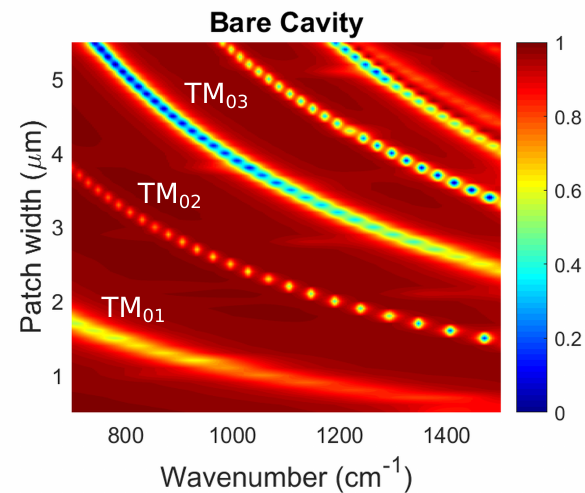}
\caption{Rigorous Coupled Wave Analysis (RCWA) simulation for a 1D metallic ribbon bare cavity, as schematized in Fig. 1(a) of the main paper.}
\label{fig:figS1}
\end{figure}

Rigorous Coupled Wave Analysis (RCWA) simulation for a 1D metallic ribbon bare cavity, as schematized in Fig. 1(a) of the main paper. The thickness of the insulator layer (GaAs) is 368.1 nm. The effective dielectric function of GaAs taking into account the interaction with phonons is in the following form:
\begin{equation*}
\epsilon_{GaAs}\,=\,\epsilon_\infty\,\left( 1+\frac{\omega_{LO}^2-\omega_{TO}^2}{\omega_{TO}^2-\omega^2-i\omega\gamma_{ph}}\right) \; ,
\end{equation*} where $\epsilon_\infty=11$ is the relative permittivity at high frequency, $\gamma_{ph}$ the damping of the phonon mode, $\omega_{LO}$ = 292 \wn\ and $\omega_{TO}$ = 268 \wn\ are the longitudinal optical phonon and transverse optical phonon frequencies, respectively \cite{Palik1998}.

The metal strip size (patch width, p in the main text) varies from 0.5 \mic\ to 5.5 \mic. The distance between the metal ribbons is kept constant at 1.5 \mic. 

Polarized light, with the electric field orthogonal to the metal ribbons, is directed onto the surface and the reflectivity is simulated for each width at the incident angle of 10 degree. For this reason the mode TM$_{02}$, which is non-radiative at normal incidence, is only barely visible. Mode TM$_{03}$ is the one with the largest contrast \cite{Balanis2016}.

\newpage
\section*{Sample HM4099 reflectivity}
\renewcommand{\thefigure}{S2}
\begin{figure}[ht]
\centering
\includegraphics[width=0.5\linewidth]{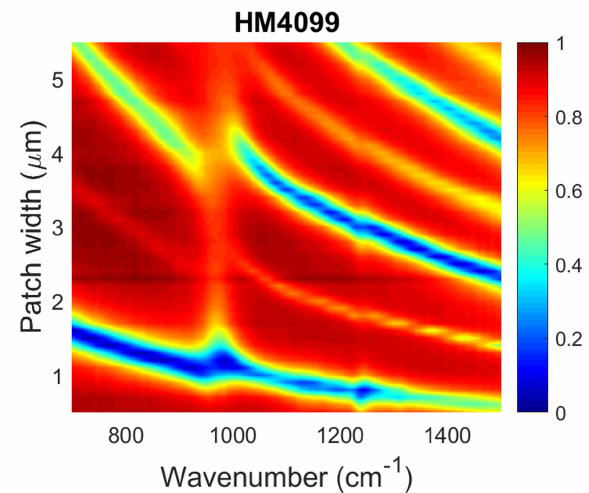}
\caption{Experimental reflectivity measurement of the doped sample HM4099 at 300 K.}
\label{fig:figS2}
\end{figure}
Experimental reflectivity measurement of the doped sample HM4099 at 300 K. The sample was nominally doped to a sheet concentration of $\text{n}_{\text{Si}} = 1.74\times 10^{12} cm^{-2}$. The bare intersubband transition is measured at 955.8 \wn.

The top metallic grating was implemented by electron beam lithography. The ribbons width varies from 0.5 \mic\ to 5.5 \mic\ and the distance between the metal stripes is 1.5 \mic. 

The reflectivity measurements were performed with a Nicolet FTIR microscope with polarized light light, resolution of 8 \wn, using a liquid nitrogen cooled MCT detector. The Rabi splitting is clearly observed for the TM$_{03}$ mode, presenting evidence of strong light-matter coupling between the cavity mode and the intersubband transition.

The modulator devices have been designed based on these measurements, in order to operate on the TM$_{03}$ mode. 

\newpage
\section*{Reflectance at 300K of sample HM4098 (N$_{Si}=6.2\times 10^{11} cm^{-2}$)}
\renewcommand{\thefigure}{S3}
\begin{figure}[h]
\centering
\includegraphics[width=0.5\linewidth]{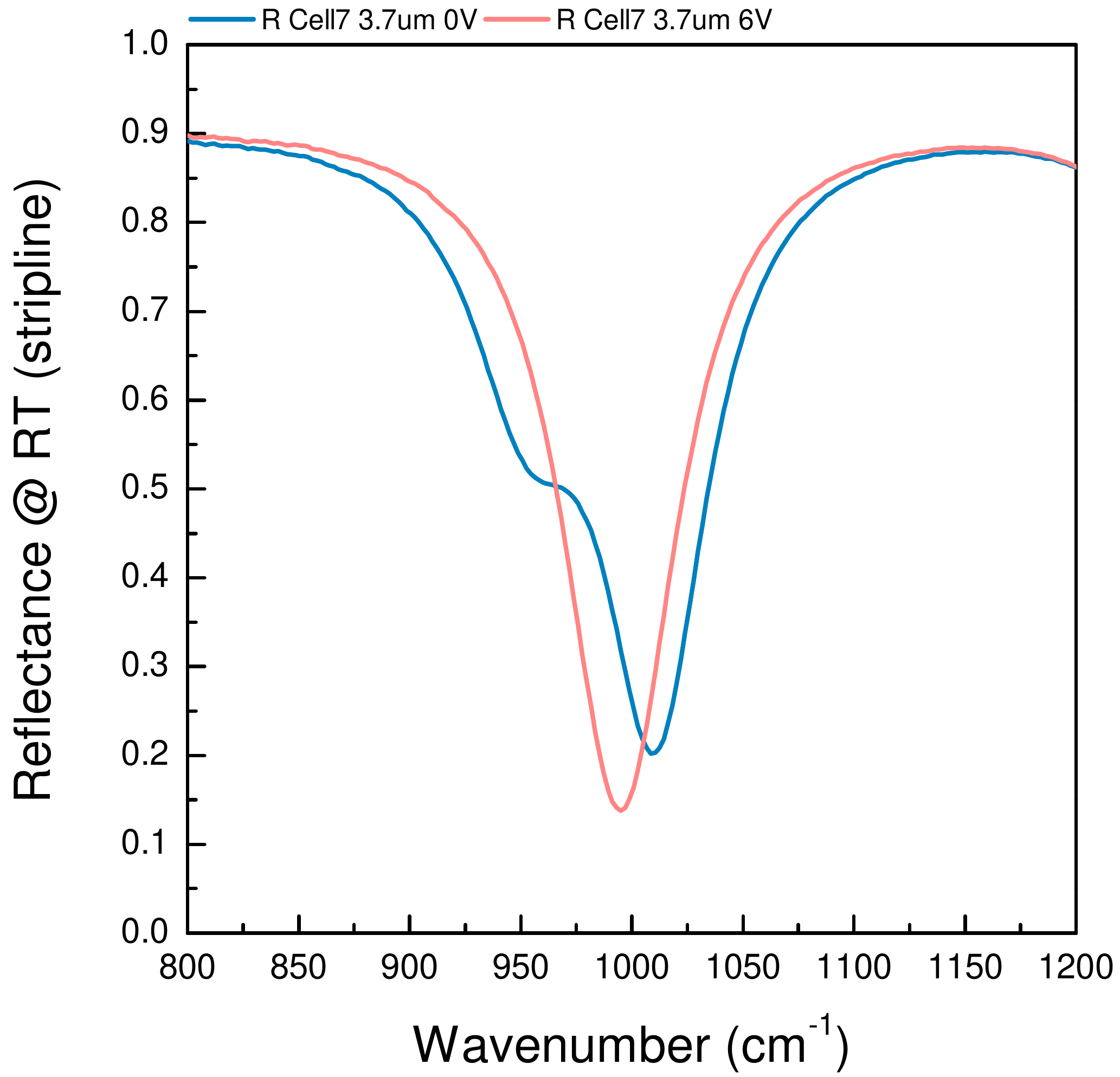}
\caption{Room temperature reflectance for HM4098 sample, with 6V bias and without any bias.}
\label{fig:figS3}
\end{figure}
We present here data on sample HM4098, that is a lower doped sample. The nominal sheet doping per QW is (N$_{Si}=6.2\times 10^{11} cm^{-2}$). 

In this case the application of a bias completely transitions the system from the strong coupling into the weak coupling regime.
However, the reduced Rabi splitting with respect to sample HM4099 yields a lower reflectivity contrast when implemented as a modulator. For this reason, we have employed sample HM4099 to implement modulators.

\printbibliography

\end{refsection}

\end{document}